\documentclass[9pt,twocolumn,twoside]{osajnl}

\journal{ol} 
\setboolean{shortarticle}{true}
\ifthenelse{\boolean{shortarticle}}{\colorlet{color2}{color2b}}{\colorlet{color2}{color2}} 

\newcommand{\bra}[1]{\langle#1\rvert} 
\newcommand{\ket}[1]{\lvert#1\rangle} 
\newcommand{\braket}[2]{ \langle #1 | #2 \rangle} 
\newcommand{\braopket}[3]{\langle {#1} | {#2} | {#3}\rangle} 

\newcommand{\me}{\mathrm{e}}
\newcommand{\mi}{\mathrm{i}}

\newcommand{\dif}{\mathrm{d}}
\newcommand{\abs}[1]{\lvert#1\rvert}

\DeclareMathOperator{\res}{\mathrm{Res}}

\newcommand{\da}{{\downarrow}} 
\newcommand{\ua}{{\uparrow}} 

\title{Few-Photon Scattering in Dispersive Waveguides with Multiple Qubits}

\author[1]{Sukru Ekin Kocabas}

\affil[1]{Department of Electrical  \& Electronics Engineering, Ko\c{c} University, Rumeli Feneri Yolu, TR34450 Sar{\i}yer, \.Istanbul, Turkey}

\affil[*]{Corresponding author: ekocabas@ku.edu.tr}

\dates{May 2015}

\ociscodes{230.7370, 270.5585, 270.4180}

\doi{}

\begin{abstract}
We extend the Krylov-subspace based time-dependent numerical simulation technique for a qubit interacting with photons in a waveguide to the multiple qubit case. We analyze photon scattering from two qubits analytically and derive expressions for the bound states in the continuum (BIC). We show how the BIC can be excited. We use the BIC in a recent Pauli-Z gate proposal involving decoherence free subspaces and obtain the gate fidelity as a function of the gate parameters. The techniques presented in the paper are useful for investigating the time evolution of quantum gates and other many-body systems with multiple quenches in the Hamiltonian.
\end{abstract}

\setboolean{displaycopyright}{false}

\begin{document}

\maketitle
\thispagestyle{fancy}

\ifthenelse{\boolean{shortarticle}}{\ifthenelse{\boolean{singlecolumn}}{\abscontentformatted}{\abscontent}}{}

Experimental advances in transmission line integrated superconducting qubits \cite{Loo2013} and atoms coupled to the near fields of waveguides \cite{Goban2015} have made it possible to build systems composed of multiple qubits, paving the way to the development of quantum information processing architectures. Investigation of such waveguide integrated multi-qubit systems in the single photon regime was primarily made through the application of the transfer matrix technique \cite{Shen2005,Tsoi2008,Zhou2008a,Chen2014}. Multiple excitation case was analyzed in \cite{Fang2014,Dai2015} under the Markovian approximation---i.e.\ assuming that the time it takes for photons to propagate between the qubits is small compared to the inverse of the atomic decay rate---and then extended to the non-Markovian case \cite{Laakso2014,Redchenko2014,Shi2015} where in both instances the waveguide dispersion, $\omega_k$, is assumed to be a linear function of the wave vector $k$. When one takes into account the dispersive nature of the modes of the waveguide, it becomes possible to form polaritonic atom-photon bound states where the photon gets trapped around the atom. Properties of the bound states were first analyzed for a single atom within a uniform photonic band gap medium \cite{John1991,Kofman1994}, and then in waveguiding geometries \cite{Calajo2016,Shi2015b}. Scattering of photonic wavepackets from bound states was analyzed in \cite{Shi2009a,Schneider2016,Kocabas2016}. Signatures of the bound states are now being probed in experiments \cite{Hood2016,Liu2016a}.

Multi-photon, multi-qubit systems are very relevant for quantum information processing, however, exact analysis of their behavior is an arduous task. Therefore, it is of interest to be able to study their dynamics independently via numerical methods. Recently, a Krylov-subspace based time evolution technique is developed to study the scattering of one- and two-photon wave packets from a waveguide embedded qubit \cite{Longo2009,Longo2010,Longo2011} where the waveguide is modeled as a series of cavities coupled to one another in a tight-binding fashion, leading to a cosine shaped dispersion relationship. Recent advances in ultrahigh-Q coupled nanocavities \cite{Notomi2008} and photonic crystal waveguides operating near their band edge \cite{Hood2016,Liu2016a} make the underlying dispersive model pertinent. In this paper, we will first generalize the Krylov-subspace based technique to the multi-qubit case, and then use the new technique to investigate various multi-qubit scenarios involving photon scattering, time evolution of a doubly excited two-qubit system, and a recent Pauli-Z gate proposal.

We begin by extending the single qubit Hamiltonian, written under the rotating-wave and dipole approximations, and introduced in \cite{Longo2009,Longo2010,Longo2011}, to the multiple qubit case. The new Hamiltonian is given by
$
H  = -J \sum_{i=1}^{L-1}( a^\dag_{i+1}a_i  + a^\dag_{i}a_{i+1})+ \sum_{s=1}^n [\frac{\Omega_s}{2} \sigma_{z_s} 
+ \bar{g}_s ( \sigma_s^+ a_{x_s} + a_{x_s}^\dag \sigma_s^- )],\nonumber
$
where $J$ is the coupling constant between neighboring cavities, $a_i$ is the annihilation operator for photons at position $i$, and $\{\Omega_s, \bar{g}_s, \sigma_{z_s}, \sigma_s^\pm\}$ are the energy level spacing, the coupling constant, the Pauli $z$ operator and the raising and lowering operator, respectively, for qubit $s$ positioned at $x_s$. There are $n$ qubits and $L$ cavities. In writing $H$ we took $\hbar=1$, assumed a normalized distance between neighboring cavities ($a=1$) and measured energies with respect to the resonant frequency of the cavities ($\omega_0=0$) \cite{Notomi2008}. With this normalization, the dispersion relation for the coupled cavity array, $\omega_0 - 2J\cos(k a)$, is transformed to $\omega_k=-2J\cos k$ with $k\in (-\pi,\pi)$. Furthermore, when doing actual calculations, we measure all energies in terms of the coupling constant ($J=1$). Therefore, distances reported in the paper are in units of $a$, energies are measured with respect to $\omega_0$ in units of $J$ and time is in terms of $1/J$. 

Our aim is to numerically calculate the time evolution of an arbitrary wave function $\ket{\psi}$. $H$ preserves the number of excitations in the system. The excitations are shared among the qubits and photons. By creating all possible combinations that lead to a fixed given excitation, we create different sectors with different states for the qubits. For instance, in the case of three excitations and two qubits, the four sectors in the system are a) three photons with both qubits in their ground states, b) two photons with the left or c) the right qubit excited, and d) one photon with both qubits excited. Once the basis sets for each sector is known, one can generate the representation of $H$ in terms of a sparse matrix. Time evolution is obtained via the operation $ \exp(-\mi H t) \ket{\psi}$. 

In \cite{Longo2009,Longo2010,Longo2011}, Kronecker product basis states that require $L^m$ elements for $m$ photons in a lattice of length $L$ are used. We use occupation basis states that require $\binom{L+m-1}{L-1}$ elements \cite{Zhang2010d}. We use the technique in \cite{Liang1995} to easily transition between different basis elements when forming $H$. There are a number of ways to numerically calculate the matrix exponent \cite{Moler2003}, we used the \texttt{Expokit} implementation \cite{Sidje1998} which comes with ready to use Matlab code.

\begin{figure}[tb]
\includegraphics{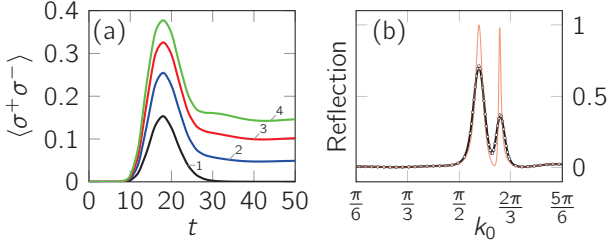}
\caption{\small (a) Excitation probability of the qubit interacting with a Gaussian pulse consisting of one through four photons, photon number shown with labels on the curves. (b) (Black line) Reflection probability of a Gaussian single photon wave packet $f_{k_0}(k)$ with center wave vector $k_0$, from two qubits separated by $R=5$ obtained via Krylov-subspace code, (white circles) reflection probability obtained via $\int \dif k \abs{r_k f_{k_0}(k)}^2$, (pink thin line) plot of $\abs{r_k}^2$ from \eqref{eq:S}.} \label{fig1}
\end{figure}

In Fig.\ \ref{fig1}(a) we show the results of scattering of a multi-photon Gaussian pulse from a single qubit, replicating a study made in \cite{Longo2010} for a lattice size of $L=99$, $\Omega=\sqrt{2}$, $\bar{g}=J=1$, pulse center wave vector $k_0=\frac{3\pi}{4}$, spatial pulse width of $5$. Our results and those in \cite{Longo2010} agree very well with each other. Whereas \cite{Longo2010} used density matrix renormalization group technique for $3$ and $4$ photons, with the new basis set, we can simulate those cases in our Krylov-subspace based code \cite{code} which we wrote using elements from the literate programming approach \cite{Knuth1992}. 

Now that we verified our numerical approach for a single qubit, let us look into the case of two qubits. We will use a resolvent based method detailed in \cite{Kocabas2016} to analyze the scattering of a photon from two qubits. The resolvent is defined as $G(z) =(z-H)^{-1}$. The Hamiltonian is written in the $k$ basis as $H=H_0 + V$ where
$
H_0 = \frac{\Omega_1}{2} \sigma_{z_1} + \frac{\Omega_2}{2} \sigma_{z_2} + \int_{-\pi}^\pi \dif k \omega_k a^\dag_k a_k, 
V = g_1 \int_{-\pi}^\pi \dif k \sigma_1^+ \me^{-\mi k R/2} a_k + g_2 \int_{-\pi}^\pi \dif k \sigma_2^+ \me^{+\mi k R/2} a_k + \text{h.c.}
$
Here "h.c." stands for hermitian conjugate, the first qubit is positioned at $x_1=-\frac{R}{2}$ and the second qubit is at $x_2=+\frac{R}{2}$, $\omega_k = -2 J \cos k$ and $2\pi g_s^2=\bar{g}_s^2$. We will have three sectors where the single excitation in the system is either in a photonic state with wave vector $k$, or in the first qubit, or in the second qubit. These three states will be shown by $\ket{k\da\da}, \ket{\ua\da}, \ket{\da\ua}$, respectively. Our aim is to calculate the scattering matrix element $\braopket{p\da\da}{S}{k\da\da}$ and compare it with numerical results from Krylov-subspace based calculations. To do so, we first write down the relationship between the $S$- and $T$-matrix elements as
$
\braopket{p\da\da}{S}{k\da\da}=\braket{p\da\da}{k\da\da}
-2\pi\mi \, \delta(\omega_p-\omega_k)\lim_{\eta\rightarrow 0^+}\braopket{p\da\da}{T(\omega_p -\frac{\Omega_1+\Omega_2}{2}+\mi\eta)}{k\da\da}.
$
The $T$-matrix elements are related to the matrix elements of the resolvent through the relationship $G(z) = G_0(z) + G_0(z) T(z) G_0(z)$ where $G_0 = (z-H_0)^{-1}$. Thus, we are tasked with finding the matrix elements of $G(z)$. We use the Lippmann--Schwinger equation for the resolvent,
$G=G_0 + G_0 V G = G_0 + G V G_0$, in conjunction with the identity operator for two qubits in the single excitation sector
$
\mathbb{1} = \ket{\ua\da}\bra{\ua\da} + \ket{\da\ua}\bra{\da\ua} + \int_{-\pi}^\pi \dif k \ket{k\da\da}\bra{k\da\da}, \nonumber
$
to derive all nine matrix elements $\braopket{p\da\da}{G}{k\da\da}$, $\braopket{p\da\da}{G}{\ua\da}$, $\braopket{p\da\da}{G}{\da\ua}$, $\braopket{\ua\da}{G}{k\da\da}$, $\braopket{\ua\da}{G}{\ua\da}$, $\braopket{\ua\da}{G}{\da\ua}$, $\braopket{\da\ua}{G}{k\da\da}$, $\braopket{\da\ua}{G}{\ua\da}$, $\braopket{\da\ua}{G}{\da\ua}$ in a manner similar to the case for a single qubit \cite{Kocabas2016}. We write down an explicit formula for $\braopket{\ua\da}{G(z)}{\ua\da}$ as
\begin{align}
\braopket{\ua\da}{G(z)}{\ua\da} & = \frac{z'-\Omega_2-g_2^2 I(z';0)}{D} \quad \text{where} \label{eq:G}\\
\begin{split}D & = [z'-\Omega_1-g_1^2 I(z';0)][z'-\Omega_2-g_2^2 I(z';0)] \\ 
& \qquad -g_1^2 g_2^2 I(z';R) I(z';-R),\end{split}\label{eq:denom}
\end{align}
with $z'=z+\frac{1}{2}(\Omega_1+\Omega_2)$. Here, the function $I(z;x)$ is defined as
$
I(z;x) = \int_{-\pi}^\pi \dif k \frac{\me^{\mi k x}}{z-\omega_k+\mi 0^+}=\frac{(-2 \pi \mi) \me^{\mi k_\star \abs{x}} }{\sqrt{4J^2-z^2}} 
= \frac{(-2 \pi \mi) \me^{\mi k_\star \abs{x}} }{2J \abs{\sin k_\star}}
$
via the use of the residue theorem for $z\in (-2J,2J)$ where $k_\star=\arccos \frac{-z}{2J}$. Through the use of the definition of $I(z;x)$ and the definition of the $S$-matrix, we can derive two-qubit transmission and reflection coefficients in terms of the ones for a single qubit where
\begin{align}
&\braopket{p\da\da}{S}{k\da\da} = t_{\mathbb{t}k} \delta(p-k) + r_k \delta(p+k) \label{eq:S}\\
&t_k = \frac{t_k^{(1)} t_k^{(2)}}{1-r_k^{(1)} r_k^{(2)} \me^{2\mi k R}} \quad
r_k = \frac{2 r_k^{(1)} r_k^{(2)} \me^{\mi k R}+r_k^{(1)} \me^{-\mi k R}+r_k^{(2)}\me^{\mi k R}}{1-r_k^{(1)} r_k^{(2)} \me^{2\mi k R}} \nonumber\\
&r_k^{(s)} = \frac{-\bar{g}_s^2}{\bar{g}_s^2 + \mi(2 J \cos k + \Omega_s)2 J \abs{\sin k}} \quad t_k^{(s)}=1+r_k^{(s)} \nonumber
\end{align}
with the qubit index $s\in\{1,2\}$ and the single qubit reflection and transmission coefficients $r_k^{(s)}, t_k^{(s)}$, respectively. These set of results can also be obtained through transfer matrix techniques \cite{Shen2005,Tsoi2008,Zhou2008a,Chen2014}, keeping in mind that the coordinate origin $x=0$ located at the midpoint of the two qubits is the input and output port plane of the two port system. In Fig.\ \ref{fig1}(b) we show the reflection probability of a single photon Gaussian pulse of spatial width 20, from two qubits with $\Omega_1=0.4$, $\Omega_2=0.8$, $\bar{g}_1=0.4$, $\bar{g}_2=0.2$ separated by 5 spatial units. We obtain the reflection coefficient numerically from Krylov-subspace based code as well as analytically via the integration of the Gaussian pulse with $r_k$. The two results agree very well with each other, providing further evidence that the code works as expected. The Gaussian pulse width in $k$ space leads to a smoothing of the $\abs{r_k}^2$ envelope. 

The Pauli-Z gate proposal that we will investigate requires us to look for bound states with energies that fall into the continuum $(-2J, 2J)$ band for propagating photons. We will refer to such states as bound states in the continuum (BIC). BIC require the presence of at least two qubits. They have been investigated in \cite{Tanaka2007,Gonzalez-Ballestero2013,Chen2016} and are different than the atom-photon bound states mentioned so far. The photonic part of the BIC is trapped between the atomic ``mirrors", can not leak out of the atomic cavity and forms a standing-wave pattern. In order to find the energies at which BIC occur, we find the location of the poles of the resolvent matrix elements from \eqref{eq:denom} for two identical qubits with $\Omega_1=\Omega_2=\Omega$ and $\bar{g}_1=\bar{g}_2=\bar{g}$. The poles are at $\Omega=z'=\omega_{k_\star}$ with $1\pm\me^{\mi k_\star R}=0$ which correspond to the even $(+)$ and odd $(-)$ solutions, respectively. For even solutions we have $k_\star=(2n_\text{e}-1)\frac{\pi}{R}$ whereas in the odd case $k_\star=2n_\text{o}\frac{\pi}{R}$ with $n_\text{e}$, $n_\text{o}$ integers. We calculate the residues of the matrix elements of $G(z)$ at the pole locations to obtain the coefficients of the elements of the bound state in continuum. Transformation from $k$-space representation to real space via $\braket{x}{k}=\frac{1}{\sqrt{2\pi}} \me^{\mi k x}$ results in the normalized bound state as
\begin{align}
\ket{\Psi^\pm_{n_\text{e,o}}} & =\mathcal{N} \left[ \ket{\ua\da}\pm\ket{\da\ua} + \sum_x (-\mi) \bar{g} \frac{\me^{\mi k_\star \abs{x+\frac{R}{2}}} \pm \me^{\mi k_\star \abs{x-\frac{R}{2}}} }{\sqrt{4J^2-\Omega^2}}\ket{x\da\da} \right]  \nonumber \\
\mathcal{N} & = \frac{1}{\sqrt{2}} \frac{1}{\sqrt{1+\frac{\bar{g}^2 R}{4J^2-\Omega^2}}},\label{eq:norm}
\end{align}
where $\mathcal{N}$ is an overall normalization constant obtained from \eqref{eq:G} via $\res(\braopket{\ua\da}{G(z')}{\ua\da},\omega_{k_\star})=\mathcal{N}^2$. The coefficients of the $\ket{\da\ua}$ and $\ket{x\da\da}$ parts of the BIC are obtained by considering $\res(\braopket{\ua\da}{G(z')}{\da\ua},\omega_{k_\star})$ and $\res(\braopket{k\da\da}{G(z')}{\ua\da},\omega_{k_\star})$. The photonic part $\braket{x\da\da}{\Psi}$ is zero for $\abs{x}>R/2$ due to the fact that $1\pm\me^{\mi k_\star R}=0$. \eqref{eq:norm} agrees with \cite{Chen2016} and extends the results obtained for a waveguide with a linear dispersion as reported in \cite{Gonzalez-Ballestero2013} to the dispersive case.

\begin{figure}[tb]
\includegraphics{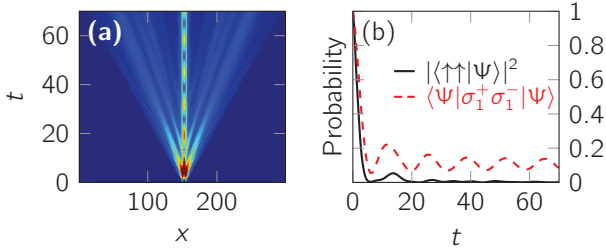}
\caption{\small Time evolution of a two qubit system initialized at $\ket{\ua\ua}$ at $t=0$ with $x_1=150$, $x_2=155$. (a) Plot of the expectation value of the number of photons as a function of space and time. (b) Plot of the probability of observing the state at $\ket{\ua\ua}$ (solid black) and the probability of having the first qubit in its excited state (dashed red) as a function of time.}\label{fig2}
\end{figure}

We used our code to evaluate the time evolution of the BIC and verified that they are indeed eigenstates of $H$ by observing that the state of the system remains unchanged as a function of time. We then evaluated the time evolution of an initially doubly excited, $\ket{\ua\ua}$, two qubit system where the separation of the qubits is $R=5$, $k_\star=\frac{\pi}{R}$, $\bar{g}=0.5$ and $\Omega_1=\Omega_2=\omega_{k_\star}$ which are parameters suitable for the formation of an even BIC. The results of the simulation are shown in Fig.\ \ref{fig2}. We see that the probability of observing $\ket{\ua\ua}$ decays down to zero whereas the probability of having the first qubit excited shows an oscillatory pattern with a corresponding bouncing photon state in between the two qubits. These results show the formation of a superposition of multiple bound states which leads to the oscillatory pattern, similar to the oscillations observed in the case of an initially excited single qubit \cite{Kocabas2016}. Excitation of the BIC for the case of a waveguide with a linear dispersion relationship was also predicted in \cite{Redchenko2014}. 

\begin{figure}[t]
\includegraphics{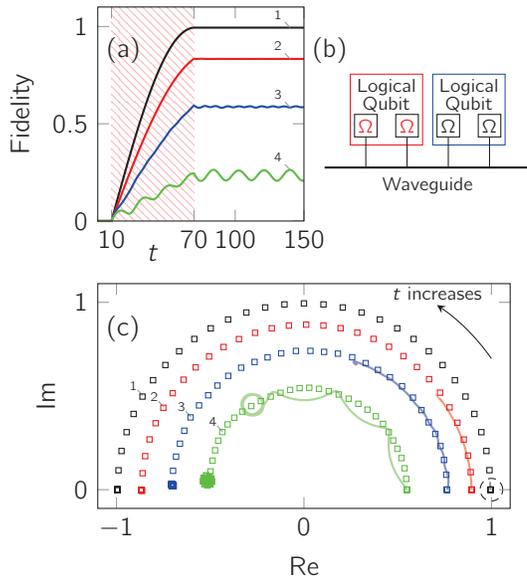}
\caption{\small (a) Fidelity as a function of $t$ in set 1 (black) for $R=4,\bar{g}=0.1,n_\text{o}=1$, set 2 (red) for $R=4,\bar{g}=0.5,n_\text{o}=1$, set 3 (blue) for  $R=7,\bar{g}=0.5,n_\text{o}=1$ and set 4 (green) for $R=7,\bar{g}=0.5,n_\text{o}=3$. (b) Sketch of the system, $H_\text{C}(t)$ is applied to the left logical qubit. (c) Evolution of $a_{01}(t)$ (square symbols) and $a_{10}(t)$ (light lines) as a function of $t$ on the complex plane for the same set of parameters in (a).}\label{fig3}
\end{figure}

We now investigate a quantum gate proposal made in \cite{Paulisch2015} for four qubits in a waveguide making use of the decoherence free subspace composed of qubit states that are antisymmetric with respect to the exchange of any two qubits. As argued in \cite{Gonzalez-Tudela2015,Gonzalez-Tudela2016,Chen2016}, the decoherence free subspace can be obtained in a one-dimensional waveguide setting through the use of the BIC. We follow the construction in \cite{Paulisch2015} and form logical qubits consisting of two neighboring physical qubits, as illustrated in Fig.\ \ref{fig3}(b). The logical qubit states are defined as $\ket{0}\equiv\ket{\da\da}$ and $\ket{1}\equiv\ket{\Psi^-_{n_\text{o}}}$. At $t=0$ we initialize the system at the superposition state $(\ket{10}+\ket{01})/\sqrt{2}$. If we only consider the qubit parts of the total wave function the initial state is given by $\frac{\mathcal{N}}{\sqrt{2}}(\ket{\ua\da\da\da}-\ket{\da\ua\da\da}+\ket{\da\da\ua\da}-\ket{\da\da\da\ua})$. We aim to incur a phase difference between the $\ket{01}$ and $\ket{10}$ states so as to test the Pauli-Z gate proposal. To do so, we apply a control Hamiltonian $H_\text{C}(t)$ where $\Delta H_{12}=\frac{\Delta}{2}(\sigma_{z_1}+\sigma_{z_2})$ is turned on for a finite duration \cite{Paulisch2015} as shown by the hatched lines in Fig.\ \ref{fig3}(a) which effectively changes the level spacing of the first two qubits from $\Omega$ to $\Omega+\Delta$ while $H_\text{C}$ is on. We can approximate the time evolution of the qubit states by considering the effects of $H_0$ and $H_\text{C}(t)$ but neglecting $V$. Note that $(H_0 + \Delta H_{12}) \ket{\ua\da\da\da}=-\Omega \ket{\ua\da\da\da}$ and similarly for $\ket{\da\ua\da\da}$. However, $(H_0 + \Delta H_{12}) \ket{\da\da\ua\da}=-(\Omega+\Delta)\ket{\da\da\ua\da}$. We see that $\Delta H_{12}$ results in an extra $\Delta$ term when the left logical qubit is in $\ket{0}$ state, but no such term exists when the left logical qubit is at $\ket{1}$ state. When we apply $H_\text{C}(t)$ for a duration $T$ and consider evolution of the initial state approximately via $\exp[\mi H_0 T] \exp[-\mi (H_0+H_\text{C})T]$, we arrive at the final state $\frac{\mathcal{N}}{\sqrt{2}}\left[\ket{\ua\da\da\da}-\ket{\da\ua\da\da}+\me^{\mi \Delta T}(\ket{\da\da\ua\da}-\ket{\da\da\da\ua})\right]$ in the interaction picture with respect to $H_0$. For $\Delta T = \pi$ the final state becomes $(\ket{10}-\ket{01})/\sqrt{2}$ and we have effectively a Pauli-(-Z) gate for the left logical qubit.

In the presence of the full Hamiltonian where qubit-photon coupling is turned on via $V$, the picture gets more complicated. We record the state of the system as a function of time in the interaction picture with respect to $H_0$ as $\frac{a_{10}(t)\ket{\underline{1}0} + a_{01}(t)\ket{0\underline{1}}}{\sqrt{2}}$ 
with $\ket{\underline{1}}\equiv\frac{1}{\sqrt{2}}(\ket{\ua\da}-\ket{\da\ua})$ denoting an ideal logic state. We also calculate the fidelity of the gate defined as $F\equiv \abs{\braket{\psi_\text{I}}{\psi(t)}}$ where the ideal final state is $\ket{\psi_\text{I}}=\frac{1}{\sqrt{2}}(\ket{\underline{1}0}-\ket{0\underline{1}})$. In Fig.\ \ref{fig3}(a) we plot $F$ as a function of time when $H_\text{C}(t)$ is turned on between $t=10$ and $t=70$ for different $R$, $\bar{g}$ and $n_\text{o}$ values. When the separation $R$ is small and $\bar{g}$ is low, we are in the Markovian regime and $F$ is close to 1. Increases in $R$, $\bar{g}$ move the system into the non-Markovian regime \cite{Gonzalez-Ballestero2013} and lower the fidelity. We can understand the reasons behind the changes in $F$ by considering the motion of $a_{10}$ and $a_{01}$ on the complex plane as time progresses. In Fig.\ \ref{fig3}(c) at $t=0$, $a_{10}=a_{01}=\sqrt{2}\mathcal{N}$ which corresponds to points on the positive real axis. As time increases, $a_{01}$ moves counter clockwise in a circular fashion towards the negative real axis. $a_{10}$ remains pinned near the positive real axis for low $\bar{g}$ as highlighted by the dashed circle in Fig.\ \ref{fig3}(c). However, increases in $\bar{g}$, $R$ or $n_\text{o}$ lead to an increased motion for $a_{10}$, lowering $F$. Furthermore, as is evident from \eqref{eq:norm}, such changes lead to a decrease in $\mathcal{N}$ which reduces $F$ as well. Oscillations in $F$ are due to bouncing trapped photon states in between the qubits. Although the Pauli-Z gate proposal in \cite{Paulisch2015} was designed assuming a Markovian model, it is interesting to note that the proposal still works, though with a lower fidelity, in the non-Markovian regime.

Before we conclude, let us briefly comment on the physical applicability of the parameters used in the paper. From the data in Fig.\ 3 of \cite{Notomi2008} we see that the value of $\kappa=2J/\omega_0$ can be varied between $4.7\times10^{-5}$ and $1.8\times10^{-3}$ by changing the distance between cavities. From Table 1 of \cite{Lodahl2015}, we see that $g=2\pi\times27$ GHz (113 {\textmu}eV) is achieved for a quantum dot in a nanobeam cavity resonant at 945 nm \cite{Ohta2011} i.e.\ $\omega_0= 2\pi\times317$ THz. Hence, for a coupled array of nanobeam cavities it seems possible to get the range $\frac{1}{2\pi}J=(7.45,285)$ GHz and $\bar{g}=\sqrt{2\pi} g \approx (0.2,9) J$ with the cavity-qubit detuning of 1 nm leading to $\abs{\Omega-\omega_0}\approx2\pi\times335 \text{ GHz}$ positioning $\Omega$ within or outside the tight-binding band of $(\omega_0-2J,\omega_0+2J)$. Typical circuit QED parameters are available in Table I of \cite{Schmidt2013}.

In conclusion, we extended the Krylov-subspace based numerical time evolution method to the multi-photon, multi-qubit case. We verified the numerical method with previously published results and analytical studies. We analyzed the bound states in the continuum and have shown how they can be excited from a doubly excited two-qubit system. We made use of the BIC in implementing a Pauli-(-Z) gate following the proposal in \cite{Paulisch2015}. We studied the properties of the gate as a function of system parameters. Our numerical method is currently using hard-wall boundary conditions requiring us to make sure that there are no reflections from the two ends of the simulation domain. Development of absorbing boundaries compatible with arbitrarily entangled many-body states would help reduce computational requirements of the simulations. Although we have not included dissipation in our formalism, introduction of a secondary waveguide to act as a reservoir is possible \cite{Rephaeli2013}. Our quantum gate is assumed to abruptly change the Hamiltonian of the system, however, better pulse shapes are conceivable \cite{Egger2014}. Our approach can be extended to simulate gates with multiple quenches, as in \cite{Nghiem2014}, to approximate arbitrarily shaped pulses. Recently, coupling between qubits and periodic waveguides was studied in \cite{Zang2015} and it would be of interest to investigate the applicability of the methods presented in this paper to such geometries. The code provided with the paper \cite{code} can be of use in simulating quantum many-body proposals utilizing bound states \cite{Douglas2015}, for studying topological order in one-dimensional waveguiding systems \cite{Grass2015} or for building modules of a quantum computation architecture \cite{Monroe2016}.

\end{document}